\documentclass[aps,prb,twocolumn,superscriptaddress,]{revtex4-2}
\usepackage{cancel}
\usepackage[normalem]{ulem}
\usepackage{bm}
\usepackage{amsmath}
\usepackage{amsthm}
\usepackage{natbib}
\usepackage{amssymb}
\usepackage{graphicx}
\makeatletter
\usepackage{times}
\usepackage{comment}
\usepackage{graphicx}
\usepackage{feynmf}
\usepackage{tabularx}
\usepackage{amsmath}
\usepackage{amstext}
\usepackage{amssymb}
\usepackage[hidelinks,colorlinks=true,urlcolor=blue,linkcolor=blue,citecolor=blue]{hyperref}
\usepackage{longtable,booktabs}
\usepackage{url}
\usepackage{subfigure}
\usepackage{amsbsy}
\usepackage{amsthm}
\usepackage{bm}
\usepackage{cleveref}
\usepackage{mathrsfs}
\usepackage{amsfonts}
\usepackage{amsbsy}
\usepackage{color}
\usepackage{physics} 

\def\brho{{\boldsymbol \brho}}

\def\nn{\nonumber}

\begin{document}
\title {Structure of the mean-field yrast spectrum of a two-component Bose gas in a ring: role of interaction asymmetry}
\author{Hui Tang}
\altaffiliation{These authors contributed equally to this work.}
\affiliation{Quantum Science Center of Guangdong-Hong Kong-Macao Greater Bay Area (Guangdong), Shenzhen 508045, China}
\affiliation{Department of Physics and State Key Laboratory of Optical Quantum Materials, University of Hong Kong, Hong Kong, China}
\author{Guan-Hua Huang}
\altaffiliation{These authors contributed equally to this work.}
\affiliation{Hefei National Laboratory, Hefei 230088, China}
\affiliation{Quantum Science Center of Guangdong-Hong Kong-Macao Greater Bay Area (Guangdong), Shenzhen 508045, China}
\author{Shizhong Zhang}
\affiliation{Department of Physics and State Key Laboratory of Optical Quantum Materials, University of Hong Kong, Hong Kong, China}
\affiliation{Hong Kong Institute of Quantum Science and Technology, University of Hong Kong, Hong Kong, China}
\author{Zhigang Wu}
\email{wuzhigang@quantumsc.cn}
\affiliation{Quantum Science Center of Guangdong-Hong Kong-Macao Greater Bay Area (Guangdong), Shenzhen 508045, China}
\author{Eugene Zaremba}
\affiliation{Department of Physics, Queen's University,
Kingston, ON,Canada}
\date{\today}
\begin{abstract}
The mean-field yrast spectrum of an SU(2)-symmetric two-component Bose gas confined to a ring geometry is known to exhibit an intricate nonanalytic structure that is absent in single-component systems. In particular, due to the interplay between the species concentration and the atomic interactions, a sequence of plane-wave states can emerge as yrast states at fractional values of the angular momentum per particle. This behavior stands in sharp contrast to the single-component case, where plane-wave states occur only at integer angular momenta. In this paper, we investigate how the structure of the yrast spectrum in a two-component Bose gas is modified by interaction asymmetry. By numerically solving the coupled Gross-Pitaevskii equations for propagating soliton states, we compute the mean-field yrast spectrum and, in particular, determine the critical curves associated with the emergence of various plane-wave yrast states. We find that both the behavior of these critical curves and the mechanisms by which plane-wave yrast states arise depend sensitively on the relative strengths of the inter- and intra-component interactions. When the intra-component interaction is weaker, the plane-wave yrast states replace soliton states through a continuous evolution, as in the SU(2)-symmetric case, although the conditions for their existence become more restrictive. In contrast, when the intra-component interaction is stronger, plane-wave yrast states may emerge by overtaking soliton states via branch crossings, and their stability is significantly enhanced. Our results have important implications for the existence and stability of persistent currents in asymmetric, two-component Bose gases. 
\end{abstract}

\maketitle
\section{Introduction}
\label{Intro}
\begin{center}
\begin{figure}[t]
     \includegraphics[width=8.6cm]{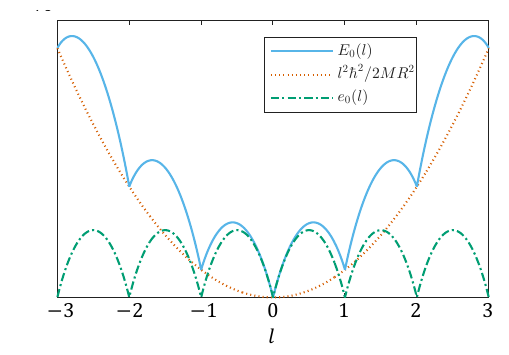}
\caption{Illustration of the yrast spectrum for a single-component superfluid in a ring of radius $R$.} 
\label{yrastill}
\end{figure}
\end{center}

The study of persistent currents in neutral superfluids confined to a multiply connected region originated from efforts to understand analogous phenomena in superconducting rings. Using arguments similar to those of Byers and Yang in their explanation of magnetic flux quantization in superconducting rings~\cite{Yang61}, Bloch showed that the occurrence of persistent currents in neutral superfluids in a ring arises from the characteristic dependence of the superfluid energy on its angular momentum per particle~\cite{Bloch73}. This energy spectrum, sometimes referred to as the yrast spectrum~\cite{Mottelson99}, takes the general form 
\begin{align}
E_0(l) = \frac{l^2\hbar^2}{2M R^2} +e_0(l),
\label{yrast0}
\end{align}
where $E_0(l)$ is the superfluid energy per particle, $M$ is the mass of the atom,  $R$ is the radius of the ring, and $l\hbar$ is the angular momentum per particle. Here $e_0(l)$ represents the internal energy which has inversion symmetry $e_0(l) = e_0(-l)$ and possesses the periodicity property $e_0(l+n)=e_0(l)$ where $n$ is an integer (see Fig.~\ref{yrastill}). These properties, along with the general assumption that $e_0(l)$ has a non-zero slope as $l$ approaches zero, imply that the yrast spectrum is not analytic at integer values of angular momentum per particle. Under suitable conditions these non-analytic points can emerge as local minima of the yrast spectrum which support persistent currents. Characteristically, the circulation of the fluid is quantized at these integer values of angular momentum per particle. This is analogous to the situation in a superconducting ring, where persistent currents are tied to the quantization of magnetic flux~\cite{Deaver61,Doll61}. 

In recent decades, quantum gases have emerged as an important platform for investigating persistent currents for two principal reasons~\cite{Ryu07,Ramanathan11,Moulder12,
Beattie13,Yakimenko2013,Wright13,Cai22,Del22,Pecci23,Polo25}. First, experimental advances in realizing multicomponent superfluids and synthetic gauge fields have substantially broadened the class of systems that may support persistent currents~\cite{KAWAGUCHI12,Lin_2016}. Second, the yrast spectrum of such systems can be determined quantitatively by mean-field theory, revealing a rich structure and enabling persistent currents beyond the conventional single-component scenario~\cite{Smyrnakis09,Anoshkin2013,Wu2013, Abad14,Wu2015,Pecci23}. Indeed, the mean-field yrast spectrum of a SU(2)-symmetric two-component Bose gas in a ring geometry is known to exhibit an intricate structure not present in single-component systems~\cite{Smyrnakis09,Anoshkin2013,Wu2013}. Although it still obeys the general form given in Eq.~(\ref{yrast0}), the yrast spectrum of this two-component system can develop a series of non-analytic points at certain fractional values of $l$, arising from the interplay between the component concentrations and atomic interactions. The states at these additional points, much like those at integer $l$, may also support persistent currents. This feature of the yrast spectrum was first uncovered using analytic soliton solutions of the coupled Gross-Pitaevskii equations for the SU(2)-symmetric system, and its implications for persistent currents were later confirmed experimentally~\cite{Beattie13}. Notably, the appearance of each additional non-analytic point at the fractional $l$ is accompanied by a transition in the condensate wave functions: a pair of soliton solutions evolve into a pair of plane-wave states. Moreover, these plane-wave states carry different winding numbers, reflecting distinct quantizations of circulation in the two components.

Most experimentally accessible two-component systems are not SU(2) symmetric, since the intra- and inter-component interaction strengths generally differ. In this case, the coupled Gross-Pitaevskii (GP) equations no longer admit analytic solutions, and it is therefore natural to ask whether the recently discovered features in the yrast spectrum persist under these more realistic conditions. Reference~\cite{Wu2015} attempted to address this question using a perturbative approach based on an assumption motivated by the SU(2)-symmetric case, namely, that the appearance of a nonanalytic point at a fractional value of the angular momentum $l$ is always accompanied by a \emph{continuous} transformation of the corresponding soliton state into a plane-wave state at that angular momentum. However, this assumption will in fact be shown not to be generally true.
Several other works~\cite{Smyrnakis2014,Sandin16, Matsushita2018,Roussou2018} have also numerically investigated the yrast spectrum of asymmetric two-component systems, but an overarching picture of the structure of the yrast spectrum has yet to emerge.  In this work, we present a comprehensive study of the mean-field yrast spectrum of asymmetric two-component Bose gases confined to a ring geometry by numerically solving the coupled GP equations. We find that the assumption adopted in Ref.~\cite{Wu2015} indeed holds for asymmetric two-component gases in which the intra-component interaction is weaker than the inter-component interaction. In the opposite regime, this assumption breaks down: plane-wave yrast states can instead emerge from the crossing of distinct branches of solutions to the GP equations, and the resulting yrast spectrum exhibits an even richer nonanalytic structure than in the SU(2)-symmetric case. In both regimes, we establish critical conditions for the emergence of plane-wave yrast states, which to a large extent determine the analytic structure of the yrast spectrum.

The rest of the paper is organized as follows. In Sec.~\ref{statement}, we review the known analytic properties of the yrast spectrum of an SU(2)-symmetric system, with particular emphasis on the critical conditions for the emergence of plane-wave yrast states. The main objective of this paper is then explicitly stated in this section: to determine the corresponding critical conditions for the asymmetric system through numerical solutions of the coupled Gross–Pitaevskii equations.  In Sec.~\ref{NM}, we develop an efficient numerical method used to solve these equations and present examples of the yrast spectrum calculated using the solutions. The structure of the yrast spectrum can be understood by the critical conditions and phase diagrams of the plane-wave yrast states which are presented in Sec.~\ref{CC}. The main results are summarized in Sec.~\ref{Conclusions}.

\section{Statement of the problem}
\label{statement}
We consider a  two-component Bose gas of $N$ atoms confined to a ring of radius $R$, where $N_A$ atoms are in hyperfine state $A$ and $N_B$ atoms in hyperfine state $B$. The atoms interact via contact interactions of strengths $U_{ss'}$ where $s$ and $s'$ label the components $A$ and $B$. We assume that $N_B <
N_A$, so that $B$ is by definition the minority component, and that 
the two components have the same mass $M$. 
Within a mean-field Gross-Pitaevskii description, the yrast spectrum (in units of $N\hbar^2/2 MR^2$) is determined by the energy functional
\begin{align}
\bar E_0 (l) =& \sum_{s }x_s\int_0^{2\pi}d\theta \left 
|\psi'_s(\theta) \right |^2 \nn \\
+&\pi\sum_{ss'}x_s x_{s'} \gamma_{ss'}\int_0^{2\pi}d\theta 
|\psi_s(\theta)|^2|\psi_{s'}(\theta)|^2,
\label{Efunc}
\end{align}
where $x_s = N_s/N$ is the component concentration and $\gamma_{ss'}=U_{ss'}NMR^2/\pi\hbar^2$ are three dimensionless interaction parameters. The condensate wave functions  $\psi_A(\theta)$ and $\psi_B(\theta)$ are the lowest energy solutions to the coupled time-independent Gross-Pitaevskii equations
\begin{align}
-\psi''_s +i \Omega\psi'_s +2\pi\sum_{s'}\gamma_{ss'}x_{s'} |\psi_{s'}|^2\psi_s 
= \mu_s \psi_s
\label{gpe0}
\end{align}
at a fixed angular momentum per particle $l\hbar$, where $\mu_s$ and $\Omega$ are (dimensionless) 
Lagrange multipliers associated, respectively, with the normalization constraints
\begin{align}
\int_0^{2\pi} d\theta |\psi_s(\theta)|^2 = 1,
\label{normalization}
\end{align}
and the angular momentum constraint
\begin{equation}
 \frac{1}{i}\sum_{s}x_s\int_0^{2\pi}d\theta 
\psi_s^*(\theta)\psi'_s(\theta) =l.
\label{agmcon}
\end{equation}
In general, the GP solutions correspond to solitons~\cite{Ohberg2001,Wu2013,Katsimiga2020} that travel around the ring at the angular velocity $\Omega$. Although the three interaction parameters may, in principle, all differ, we focus on the experimentally relevant case in which the intra-component interactions are equal but differ from the inter-component interaction, i.e., $\gamma_{AA}=\gamma_{BB}\neq \gamma_{AB}$. We therefore set 
\begin{align}
\gamma_{AB}&=\gamma; \nn \\
\gamma_{AA}&=\gamma_{BB}=(1+\kappa)\gamma,
\end{align}
 where the dimensionless parameter $\kappa$ characterizes the interaction asymmetry. In the special case of $\kappa = 0$, the system possesses the SU(2) symmetry and Eq.~(\ref{gpe0}) can be solved analytically in terms of Jacobi elliptic functions~\cite{Wu2013}.

\begin{center}
\begin{figure}[t]
     \includegraphics[width=8.6cm]{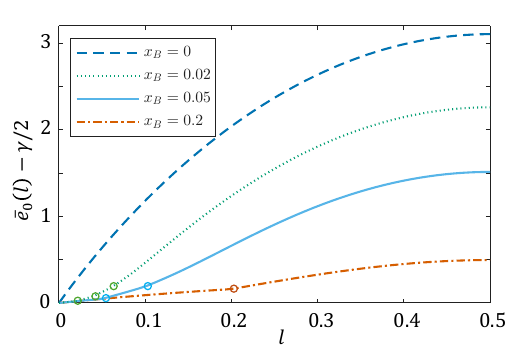}
\caption{The internal part of the mean-field yrast spectrum of an SU(2)-symmetric system, plotted in the fundamental range $0 < l \le 1/2$ for $\gamma = 100$. The circles mark the points at which the derivative of the spectrum is discontinuous and at which the corresponding yrast state is a plane-wave state.} 
\label{yrastsu2}
\end{figure}
\end{center}

\begin{center}
\begin{figure}[t]
     \includegraphics[width=8.6cm]{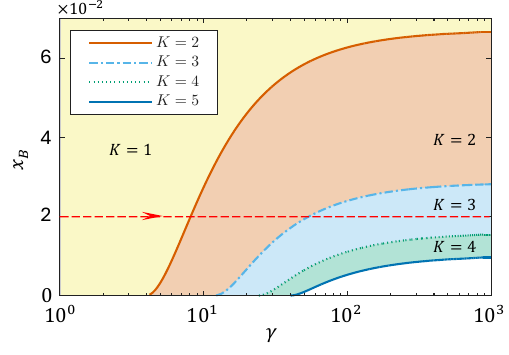}
\caption{Critical curves for plane-wave yrast states $(\phi_0,\phi_K)$ with $K=2,3,4$ and the corresponding phase diagram for the SU(2)-symmetric system. Note that the $\gamma$ axis is plotted on a logarithmic scale.} 
\label{ccsu2}
\end{figure}
\end{center}

The purpose of our paper is to determine the analytic structure of the mean-field yrast spectrum $\bar E_0(l)$ as a function of $l$ for the asymmetric system. To phrase our objective more precisely, we first review the known properties of the yrast spectrum and the associated condensate wave functions in the SU(2)-symmetric system. Because of the general form in Eq.~(\ref{yrast0}), we need only to consider the internal energy (in units of $N\hbar^2/2 MR^2$)  
\begin{align}
\bar e_0(l) \equiv \bar E_0(l) - l^2
\end{align}
in the fundamental
range $0< l\le 1/2 $.  Unlike the single-component gas, for which the yrast spectrum in this range is a smooth curve, the derivative of the spectrum in the SU(2)-symmetric two-component system exhibits discontinuities at $l=kx_B$, where $k = 1,2,..., K$ (see Fig.~\ref{yrastsu2}). A derivative discontinuity at a given value $l= Kx_B$ appears when the parameters $(\gamma, x_B)$ lie within the region of the $\gamma$-$x_B$ plane bounded by the corresponding critical curve $x_{B}(\gamma,K)$ and the $x_B = 0$ axis. The critical curve for $K = 1$ is simply the line $x_B = 1/2$, implying that the discontinuity at $l=x_B$ is always present. The critical curves for $K = 2,3,4$ are illustrated in Fig.~\ref{ccsu2}; each curve approaches the asymptotic value $\big(1-\sqrt{1-1/K^2}\big )/2$ as $\gamma\rightarrow \infty$.  Thus, the spectrum contains a total of $K$ derivative discontinuities when the coordinate $(\gamma, x_B)$ lies within a region 
bounded by the two critical curves $x_{B}(\gamma,K)$ and $x_{B}(\gamma,K+1)$~\cite{Wu2015}, as shown in Fig.~\ref{ccsu2}. In view of this diagram, the variation of the yrast-spectrum structure with $x_B$, shown in Fig.~\ref{yrastsu2}, is readily understood.

The non-analyticity at these points indicates that the corresponding condensate wave functions are special compared with those at arbitrary values of $l$. Indeed, at the non-analytic $l = kx_B$ points the condensate wave function takes a plane-wave form  $(\psi_A,\psi_B)= ( \phi_0,\phi_k)$ where 
\begin{align}
\phi_k=e^{i k\theta}/\sqrt{2\pi},
\end{align}
whereas the yrast state is generally a soliton state at other fractional values of $l$. Consequently, the critical curves $x_B(\gamma,K)$ in Fig.~\ref{ccsu2} can also be interpreted as delineating the phases in which different numbers of plane-wave yrast states appear. It is interesting to see how these plane-wave yrast states emerge from soliton states as the interaction strength changes. For example, consider a system with $x_B = 0.02$ and imagine increasing $\gamma$ from a small value along the dashed line shown in Fig.~\ref{ccsu2}). Since this value of $x_B$ lies within the range
 \begin{align*}
  \frac{1}{2}\left [ 1-\sqrt{1-1/(K+1)^2}\right ] <  x_B < \frac{1}{2}\left [1-\sqrt{1-1/K^2}\right ]
 \end{align*}
 with $K=3$, the yrast spectrum has at most three derivative discontinuities in the fundamental range. As noted earlier, the plane wave state $(\phi_0,\phi_1)$ is always the yrast state at $l = x_B$, independent of $\gamma$. The yrast spectrum develops a second derivative discontinuity at $l = 2x_B$ once $\gamma \gtrsim  8.1$, a value determined by the intersection of the line $x_B = 0.02$ with the critical curve $x_B(\gamma,2)$. As $\gamma$ approaches this critical value from below, the soliton state at $l = 2x_B$ continuously evolves into the plane-wave state $(\phi_0,\phi_2)$. As $\gamma$ continues to increase beyond this critical value this plane wave persists as the yrast state at $l = 2x_B$. Finally, a third and final plane-wave yrast state $(\phi_0,\phi_3)$ emerges at $l=3x_B$ when $\gamma \gtrsim 54.2$, as determined by the intersection of the same line with the critical curve $x_B(\gamma,3)$.

The above summary shows that, for the SU(2)-symmetric system, the yrast spectrum in the fundamental range consists of smooth soliton branches joined by plane-wave states $(\phi_0,\phi_k)$ at fractional values $l = kx_B$, where each junction appears as a nonanalytic point. Owing to the periodicity and the inversion symmetry of the internal energy, the appearance of the plane-wave yrast state $(\phi_0,\phi_k)$ necessarily implies the simultaneous appearance of the states $(\phi_\mu,\phi_{\mu\pm k})$ where $\mu = 1,2,\cdots$.  In other words, the analytic structure of the yrast spectrum is fundamentally determined by the emergence of these plane-wave yrast states. We can now state our problem more concisely: for the asymmetric system, which plane-wave states can become yrast states, and under what conditions? Or, equivalently, what are the critical curves and phase diagram of the plane-wave yrast states for the asymmetric system? To address these questions we obtain numerical solutions of the GP equations. In the process, we will determine whether or not plane-wave yrast states emerge from soliton states in a continuous manner.

\section{Numerical method}
\label{NM}
We employ the method of imaginary time propagation to obtain solutions to the coupled GP equations in Eq.~(\ref{gpe0}) for $0\le l \le 1/2$, which are then substituted in Eq.~(\ref{Efunc}) to calculate the yrast spectrum. Our method differs from that employed in Ref.~\cite{Sandin16}. The key challenge in solving Eq.~(\ref{gpe0}) is to implement the angular momentum constraint in Eq.~(\ref{agmcon}), or equivalently, to determine the Lagrange multiplier $\Omega$ in terms of $l$ and the condensate wave functions. For this purpose, we make use of the modulus-phase representation
\begin{center}
\begin{figure}[ht]
     \includegraphics[width=8.6cm]{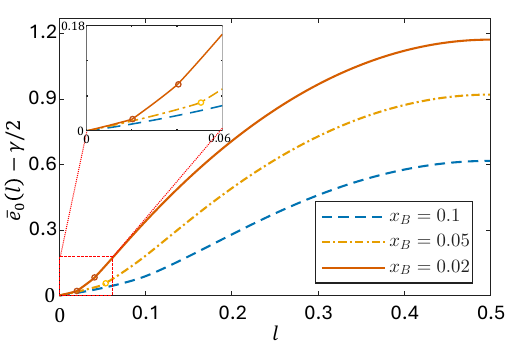}
\caption{The internal part of the mean-field yrast spectrum of an asymmetric system with $\kappa = -0.1$, plotted in the fundamental range $0 < l \le 1/2$ for $\gamma = 20$. The circles indicate the locations at which the derivative of the spectrum is discontinuous.} 
\label{yrastkm}
\end{figure}
\end{center}
\begin{equation}
\psi_s(\theta)=\sqrt{\rho_s(\theta)}e^{i\varphi_s(\theta)}.
\label{2para}
\end{equation} 
\begin{center}
\begin{figure}[t]
     \includegraphics[width=8.6cm]{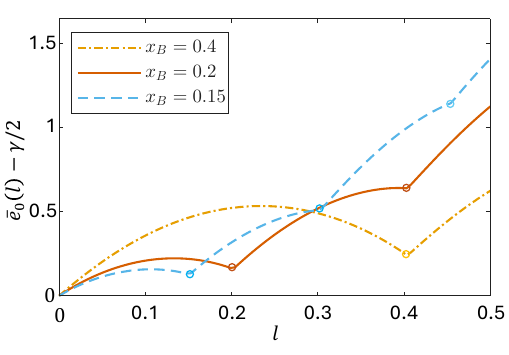}
\caption{The internal part of the mean-field yrast spectrum of an asymmetric system with $\kappa = 1$, plotted in the fundamental range $0 < l \le 1/2$ for $\gamma = 20$. The circles mark the points at which the derivative of the spectrum is discontinuous and at which the corresponding yrast state is a plane-wave state.} 
\label{yrastkp}
\end{figure}
\end{center}
The single-valuedness of the wave function $\psi_s(\theta)$ implies the following boundary conditions 
\begin{align}
\label{2bdc1}
\rho_s(\theta+2\pi)-\rho_s(\theta)&=0 \\
\varphi_s(\theta+2\pi)-\varphi_s(\theta)&=2\pi J_s, \quad J_s=0,\pm 
1,\pm 2,\cdots
\label{2bdc2}
\end{align}
where the integers $J_s$ are referred to as phase winding numbers. Substituting Eq.~(\ref{2para}) into Eq.~(\ref{gpe0}), one obtains
\begin{equation}
-\frac{\rho''_s}{2\rho_s}+\frac{\rho'_s}{4\rho^2_s}+\left (\varphi'_s -\Omega\right 
)\varphi'_s+2\pi\sum_{s'}\gamma_{ss'}x_{s'} \rho_{s'} =\mu_s
\label{ampeqa}
\end{equation}
and
\begin{equation}
\sqrt{\rho_s}\varphi''_s+\left (2\varphi'_s-\Omega\right )\left( 
\sqrt{\rho_s}\right )'=0.
\label{pheqa}
\end{equation}
Equation~(\ref{pheqa}) can be solved as 
\begin{equation}
\varphi'_s(\theta)=\frac{W_s}{2\rho_s}+\frac{\Omega}{2},
\label{pheq_sola}
\end{equation}
where $W_s$ is an integration constant. Integrating the above equation and using Eq.~(\ref{2bdc2}), we find 
\begin{align}
W_s = \frac{4\pi J_s - 2\pi \Omega}{I_s},
\label{Ws}
\end{align}
where
\begin{align}
I_s \equiv \int_0^{2\pi} d\theta \frac{1}{\rho_s(\theta)} .
\end{align}
Using Eq.~(\ref{agmcon}) and Eq.~(\ref{pheq_sola}), we 
find that the angular momentum per particle can be written as 
\begin{equation}
l=x_Al_A+x_Bl_B,
\label{agconsab}
\end{equation}
where 
\begin{equation}
l_s={\pi W_s}+\frac{\Omega}{2}
\label{agms}
\end{equation}
is the angular momentum per particle of each species. Substituting Eq.~(\ref{Ws}) in Eq.~(\ref{agms}) and using Eq.~(\ref{agconsab}) we finally arrive at
\begin{align}
\Omega = \frac{l - 4\pi^2(x_AJ_A/I_A+x_BJ_B/I_B)}{1/2 - 2\pi^2 (x_A/I_A + x_B/I_B)}.
\label{Omegaequ}
\end{align}
We have thus expressed $\Omega$ in terms of $l$, as well as the modulus and the phase winding numbers of the condensate wave functions. In the imaginary time propagation of Eq.~(\ref{gpe0}) at a specific $l$, the wave functions obtained after each iteration are used to extract the parameters $I_s$ and the phase winding numbers $J_s$; these are used in Eq.~(\ref{Omegaequ})  to evaluate $\Omega$, and the resulting $\Omega$ is then fed back into the next iteration.  Such a procedure leads to a very efficient algorithm for solving the coupled GP equations. As a benchmark, our numerical solutions for the SU(2)-symmetric case accurately reproduce the yrast spectrum obtained from the analytic soliton solutions in Fig.~\ref{yrastsu2}, as well as the critical curves and phase diagram in Fig.~\ref{ccsu2}. 

\begin{center}
\begin{figure}[t]
     \includegraphics[width=8.6cm]{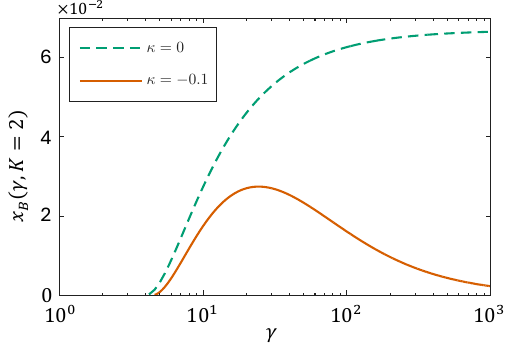}
\caption{The critical curves for the plane-wave yrast state $(\phi_0,\phi_2)$ at $\kappa = -0.1$ (solid line) and $\kappa = 0$ (dashed line). } 
\label{cckn}
\end{figure}
\end{center}
\begin{center}
\begin{figure}[t]
     \includegraphics[width=8.6cm]{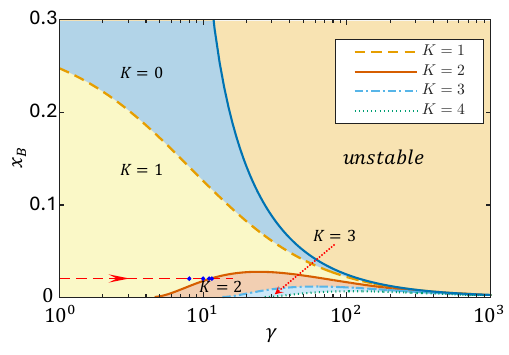}
\caption{The critical curves for the emergence of plane-wave yrast states and the corresponding phase diagram for $\kappa = -0.1$. Here the value of $K$ indicates the total number of plane-wave yrast states in that phase.} 
\label{pdkn}
\end{figure}
\end{center}

In Fig.~\ref{yrastkm}, we present examples of the internal yrast spectrum calculated for several values of $x_B$ at $\kappa = -0.1$ and $\gamma = 20$. We observe that the spectrum exhibits no derivative discontinuity for $x_B = 0.1$, a single discontinuity at $l = 0.05$ for $x_B = 0.05$, and two discontinuities at $l = 0.02$ and $l = 0.04$ for $x_B = 0.02$. Similar results for $\kappa = 1$ and $\gamma = 20$ are shown in Fig.~\ref{yrastkp}, where the spectrum exhibits one derivative discontinuity at $l = 0.4$ for $x_B = 0.4$, two discontinuities at $l = 0.2$ and $l = 0.4$ for $x_B = 0.2$, and three discontinuities at $l = 0.15$, $0.3$, and $0.45$ for $x_B = 0.15$. At all of these discontinuities, located at $l = k x_B$, the condensate wave functions correspond to the plane-wave states $(\phi_0,\phi_k)$. All of these features of the yrast spectrum will become transparent once the critical conditions for the emergence of plane-wave yrast states, discussed below, are established.
Lastly, we observe in Fig.~\ref{yrastkp} that the slope of the spectra at $l = 0.5$ is finite, in contrast to the vanishing slope for systems with $\kappa \le 0$. In light of the inversion symmetry, this implies that the derivative at this angular momentum is discontinuous. In the following we will examine the underlying origin of this phenomenon and assess whether it is a general feature of asymmetric systems with $\kappa > 0$. 

\section{Critical conditions and phase diagrams}
\label{CC}
\begin{figure*}[t]
\begin{centering}
\includegraphics[width=17.4cm]{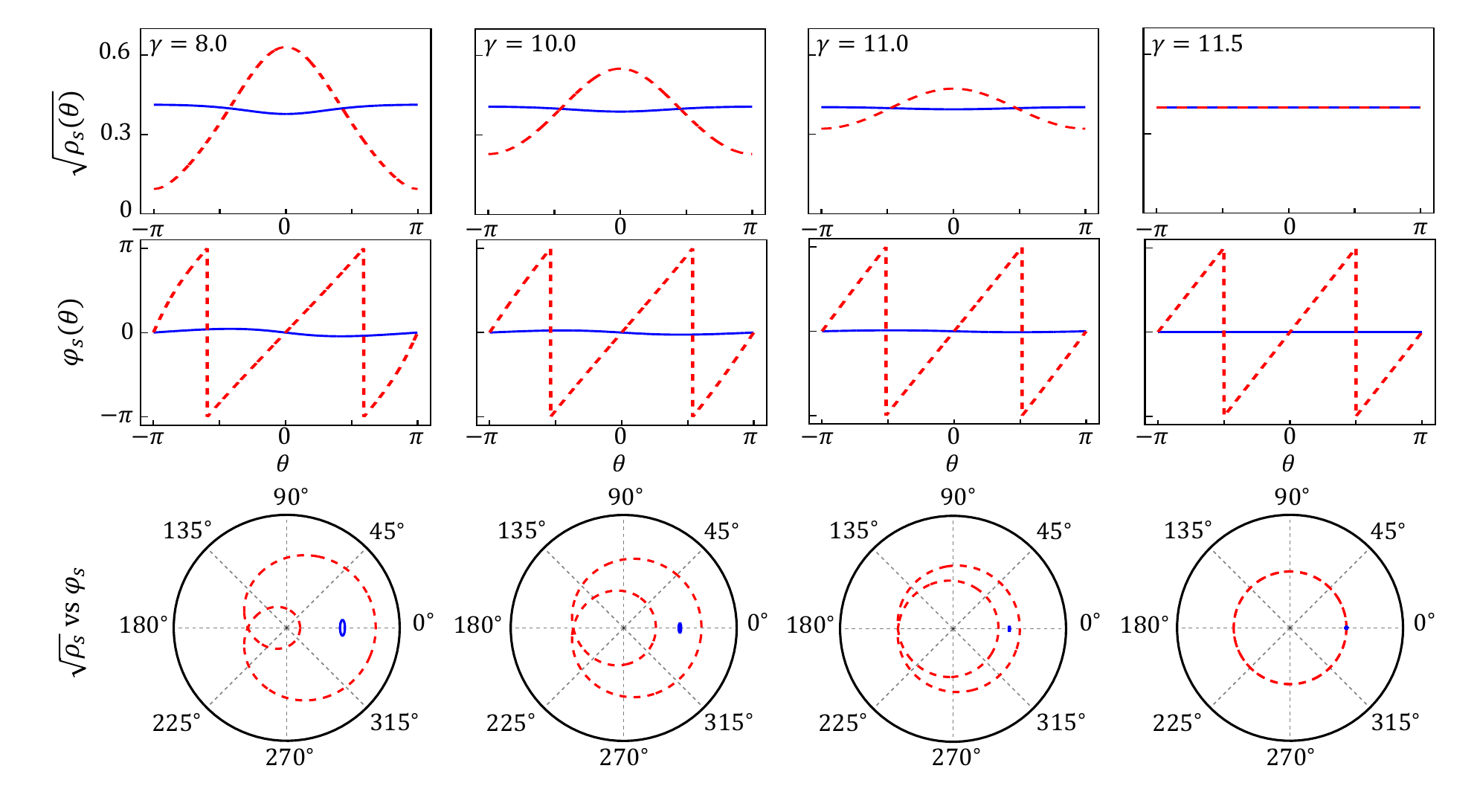}
\par\end{centering}
\caption{The upper and middle panels, respectively, show the amplitude and phase (modulo $2\pi$) of the condensate wave function at $l = 2x_B$ with $x_B = 0.02$ and $\kappa = -0.1$ for $\gamma = 8,10,11$ and $11.5$ (see the dots on the dashed line with arrow in Fig.~\ref{pdkn}). The continuous transition from the soliton state to the plane-wave state $(\phi_0,\phi_2)$ as $\gamma$ crosses the critical curve of $K=2$ is most clearly illustrated by a plot of the amplitude as a function of the phase in the polar coordinate, displayed in the bottom panel. {The solid (blue) and the dashed (red) lines represent the A and B components, respectively. }}
\label{wf_km}
\end{figure*}
 To determine the critical curve $x_B(\gamma,K)$ for the emergence of the $(\phi_0,\phi_K)$ yrast state, we can restrict the possible $K$ to positive integers because of the inversion symmetry of the yrast spectrum. It is clear that these curves also depend on the parameter $\kappa$ characterizing the interaction asymmetry.  We now present a simple argument showing that the behavior of these critical curves differ qualitatively for $\kappa < 0$ and $\kappa > 0$. From Eq.~(\ref{Efunc}) we find that the energy of the plane-wave state $(\phi_0,\phi_k)$ is given by
\begin{align}
\bar E_0 = x_B k^2    + \frac{1}{2}\left [ \gamma + \kappa \gamma (x_A^2 + x_B^2)\right ].
\end{align}
The energy difference $\delta \bar E_0$ between an arbitrary state $(\psi_A,\psi_B)$ in the same $l = kx_B$ manifold and this plane-wave state naturally separates into kinetic and interaction contributions, i.e., 
\begin{align}
\delta \bar E_0(\kappa) = \delta \bar E_{\rm kin} + \delta \bar E_{\rm int}(\kappa),
\label{deltaE0}
\end{align}
where the kinetic energy difference does not depend on $\kappa$ and the interaction energy difference is given explicitly by 
\begin{align}
\delta \bar E_{\rm int}(\kappa) &=\pi \kappa\gamma  \left[ x_A^2\int_0^{2\pi} d\theta |\delta \rho_A|^2 +  x_B^2\int_0^{2\pi} d\theta |\delta \rho_B|^2 \right ]\nn \\
 & +   \pi \gamma \int_0^{2\pi} d\theta |\delta \rho |^2
 \label{deltaEint}
\end{align}
with $\delta\rho_s(\theta) = \rho_s(\theta) - 1/2\pi$ and $\delta \rho(\theta) = x_A\delta\rho_A(\theta) + x_B\delta \rho_B(\theta)$. 
Now suppose that the plane-wave state $(\phi_0,\phi_k)$ is an yrast state at $\kappa =0$ for some  parameter set $(\gamma,x_B)$, meaning that it has the lowest energy in the $l = kx_B$ manifold, i.e., 
\begin{align}
\delta \bar E_0(\kappa =0) = \delta \bar E_{\rm kin}  + \delta \bar E_{\rm int}(\kappa =0 )  > 0.
\end{align}
Under the same parameters, this plane-wave state is then guaranteed to remain an yrast state for $\kappa > 0$, since 
\begin{align}
\delta \bar E_0(\kappa >0)> \delta \bar E_0(\kappa =0). 
\end{align}
Conversely, this state may cease to be an yrast state when $\kappa < 0$, since 
\begin{align}
\delta \bar E_0(\kappa <0)< \delta \bar E_0(\kappa =0),
\end{align}
and decreases with decreasing $\kappa$.
The above analysis shows that an asymmetric interaction with $\kappa>0$ stabilizes the plane-wave state as an yrast state, whereas $\kappa< 0$ tends to destabilize it. Given the saturating behavior of the critical curves at $\kappa = 0$ shown in Fig.~\ref{ccsu2}, we expect the curves for $\kappa< 0$ to move down and away from the asymptotic lines, while those for $\kappa>0$ to cross and move above these lines. 
\subsection{$\kappa < 0$}
As an example, we show in Fig.~\ref{cckn} the critical curve for the plane-wave yrast state $(\phi_0,\phi_2)$ at $\kappa = -0.1$, together with the corresponding curves for $\kappa =0$. The curve at $\kappa = -0.1$ behaves exactly as anticipated from the preceding argument. In particular, the critical values of $x_B$ approach zero at large $\gamma$.  Since the plane-wave state $(\phi_0,\phi_k)$ becomes an yrast state in the region bounded by its critical curve and the $x_B=0$ axis, this implies that this state, after becoming an yrast state, will eventually cease to be an yrast state at sufficiently large $\gamma$, regardless of the value of $x_B$. Such a behavior can also be understood from Eqs.~(\ref{deltaE0}) and (\ref{deltaEint}), which show that $\delta \bar E_0(\kappa<0)$ decreases without bound as $\gamma$ increases and will ultimately become negative. In other words, for a fixed negative value of $\kappa$, increasing $\gamma$ tends to destabilize the plane-wave state as an yrast state. This is most clearly illustrated by the $(\phi_0,\phi_1)$ state: it is always an yrast state at $\kappa = 0$, yet for $\kappa < 0$ it ceases to be one once $\gamma$ exceeds a certain threshold (see Fig.~\ref{pdkn}). {  In fact, the entire system becomes unstable once $\gamma$ exceeds values set by the dynamical instability condition~\cite{Anoshkin2013}
\begin{align}
\left[x_A\tilde\gamma + {1}/{2}\right]\left[x_B\tilde\gamma + {1}/{2}\right]< x_Ax_B\gamma^2,
\label{inscon}
\end{align}
where $\tilde \gamma = (1+\kappa)\gamma$. 
This condition can be obtained by analyzing the yrast spectrum in the vicinity of $l=0$.  In the $l\rightarrow 0$ limit, the yrast spectrum is essentially the lower branch of the Bogoliugov excitation spectrum, which can be obtained by Bogoliubov transformation as~\cite{Anoshkin2013}
\begin{align*}
\bar E_0(l)  =   l \sqrt{(Nl)^2 +\tilde\gamma -\sqrt{(x_A-x_B)^2 {\tilde \gamma}^2 + 4x_Ax_B\gamma^2} }.
\end{align*}
Since the total angular momentum $L = Nl\hbar$ is quantized, the smallest finite angular momentum is $L = \hbar$, which gives the lowest excitation energy as \begin{align}
\bar E_0(1/N) = \frac{1}{N}\sqrt{1+\tilde\gamma -\sqrt{(x_A-x_B)^2 {\tilde \gamma}^2 + 4x_Ax_B\gamma^2}}.
\end{align} If this energy becomes imaginary, i.e., if 
\begin{align}
1+\tilde\gamma -\sqrt{(x_A-x_B)^2 {\tilde \gamma}^2 + 4x_Ax_B\gamma^2} < 0,
\end{align}
then it signals an instability of the system. The above inequality is in fact equivalent to the dynamical instability condition in Eq.~(\ref{inscon}).}

The instability curve together with a partial phase diagram of the plane-wave yrast states is shown in Fig.~\ref{pdkn} for $\kappa = -0.1$. It is not feasible to display all the phases since phases hosting an increasing number of plane-wave yrast states emerge at progressively smaller values of $x_B$ and increasingly larger values of $\gamma$. The nonanalytic structure of the yrast spectrum shown in Fig.~\ref{yrastkm} can now be understood in light of the corresponding phase diagram. Specifically, for a given set of parameters $(x_B,\gamma)$, the number of derivative discontinuities in the spectrum associated with plane-wave yrast states is predicated by the location of $(x_B,\gamma)$ within the phase diagram. For instance, the yrast spectrum for $(x_B,\gamma) = (0.02, 20)$ has two derivative discontinuities at $l = x_B$ and $2x_B$ because the parameter set lies in the $K = 2$ phase.

Interestingly, we find that the critical curves for $\kappa<0$ obtained from numerical solutions of the coupled GP equations agree with those given in Ref.~\cite{Wu2015} using a perturbative approach. { In this approach, one begins at a sufficiently large $\gamma$ that a specific plane-wave state is guaranteed to be an yrast state at angular momentum $l = kx_B$. In this regime, the local structure of the yrast spectrum in the vicinity of this plane-wave state can be determined perturbatively, and it exhibits a cusp at $l = kx_B$ (see Fig.~\ref{yrastkp} for an example). As $\gamma$ is gradually decreased, this cusp becomes progressively less pronounced and eventually smooths out and its disappearance defines the critical value of $\gamma$. This approach, however, presupposes that at $l = kx_B$ the plane-wave state transitions continuously into a soliton state as $\gamma$ is lowered.}   The agreement between the analytic approach and our numerical simulation here then implies that the assumption is correct for $\kappa <0$. This is verified by our numerical calculations. As a representative example, we fix $x_B = 0.02$ and increase $\gamma$. As shown in the phase diagram in Fig.~\ref{cckn}, the plane-wave state $(\phi_0,\phi_2)$ becomes an yrast state once $\gamma$ exceeds the critical value $ \gamma_c \simeq 11.2$. In Fig.~\ref{wf_km}, we plot the condensate wave functions at $l = 2 x_B$ for $\gamma = 8$, $10$, $11$, and $11.5$. These results clearly demonstrate a continuous evolution from soliton states to the plane-wave state $(\phi_0,\phi_2)$ as $\gamma$ approaches and crosses the critical value.

\subsection{$\kappa > 0$}
\begin{center}
\begin{figure}[t]
     \includegraphics[width=8.6cm]{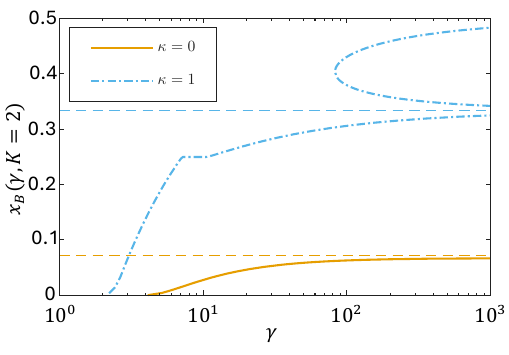}
\caption{The critical curves for the plane-wave yrast state $(\phi_0,\phi_2)$ at $\kappa = 0$ (solid line) and $\kappa = 1$ (dashed line).} 
\label{cckpo}
\end{figure}
\end{center}
\begin{center}
\begin{figure}[t]
     \includegraphics[width=8.6cm]{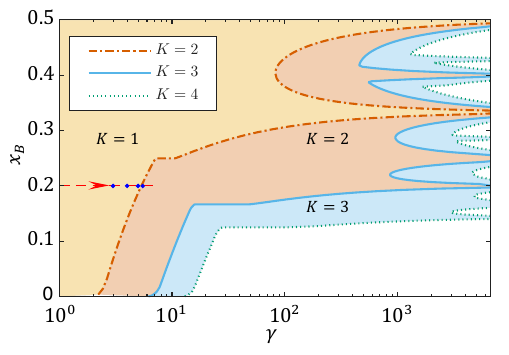}
\caption{The critical curves for plane-wave yrast states $(\phi_0,\phi_k)$ with $k=2,3,4$ and the corresponding phase diagram for $\kappa = 1$.} 
\label{cckp}
\end{figure}
\end{center}

\begin{center}
\begin{figure}[bh]
     \includegraphics[width=8.6cm]{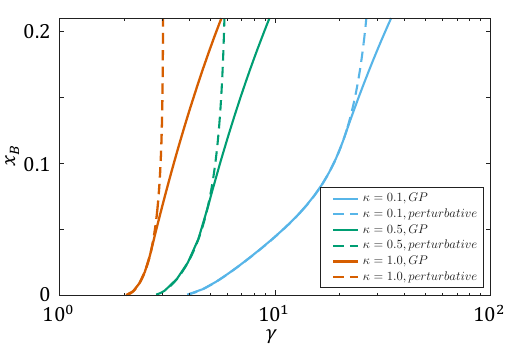}
\caption{Comparison of the $K=2$ critical curves determined by the perturbative approach (dashed lines) and by the GP analysis (solid lines) for $\kappa = 0.1, 0.5$ and $1$.} 
\label{cccomparison}
\end{figure}
\end{center}

\begin{figure*}[t]
\begin{centering}
\includegraphics[width=17.4cm]{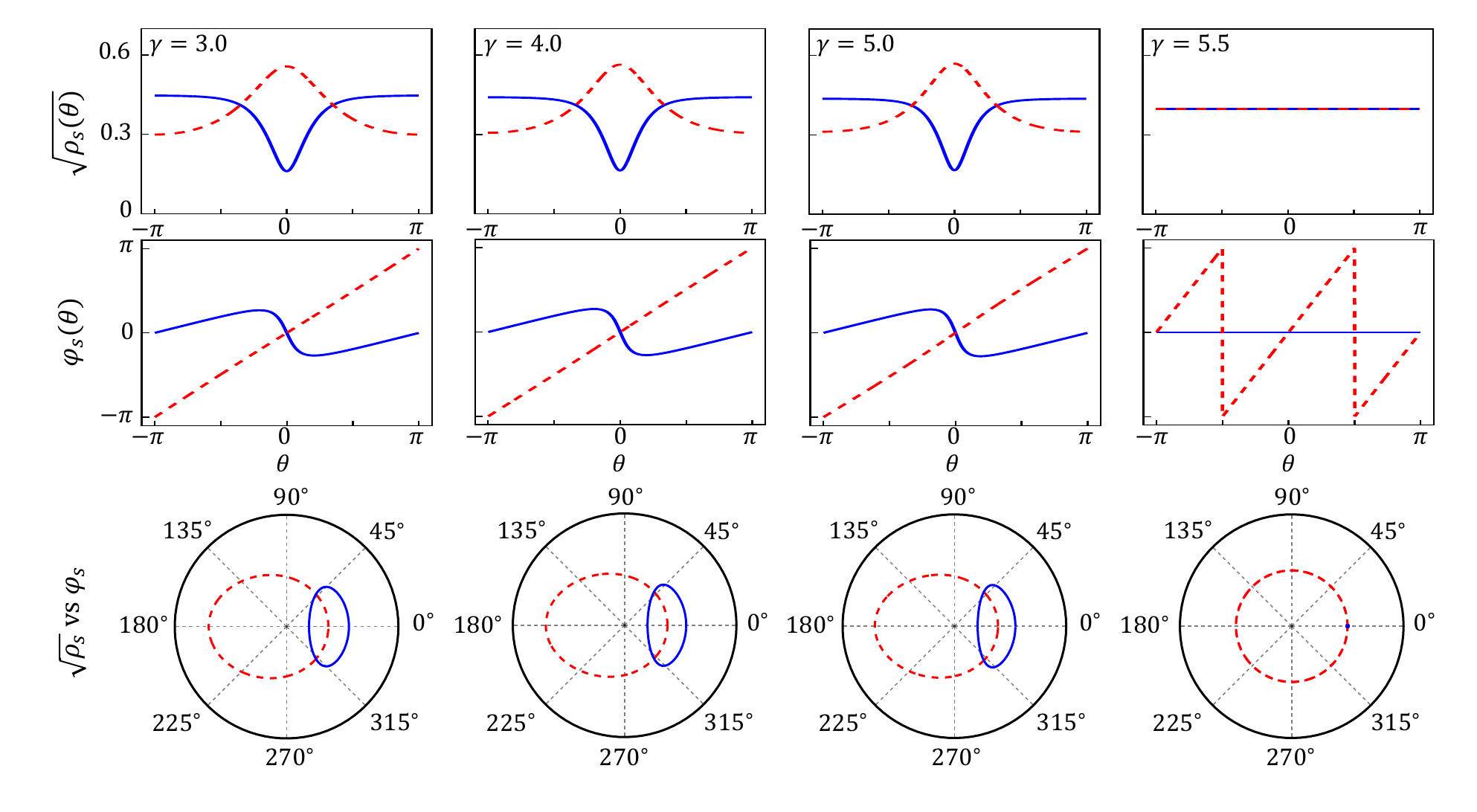}
\par\end{centering}
\caption{The upper and middle panels, respectively, show the amplitude and phase of the condensate wave function at $l = 2x_B$ with $x_B = 0.2$ and $\kappa = 1$ for  $\gamma = 3,4,5$ and $5.5$ (see the dots on the dashed line with arrow in Fig.~\ref{cckp}). The bottom panel plots the amplitude of the condensate wave function against the phase in the polar coordinate. { The solid (blue) and the dashed (red) lines represent the A and B components, respectively. } }
\label{wf_kp}
\end{figure*}
Since $(\phi_0,\phi_1)$ is always an yrast state for $\kappa = 0$, the preceding analysis implies that it remains an yrast state for $\kappa > 0$ as well. Turning to the critical curve of the plane-wave state $(\phi_0,\phi_2)$ for positive $\kappa$, we find that it indeed crosses the $K=2$ asymptote for $\kappa = 0$ as $\gamma$ increases, as illustrated in Fig.~\ref{cckpo}. Interestingly, once the asymptotic line at $\kappa = 0$ is crossed, additional structure emerges. The critical value of $x_B$ continues to increase with $\gamma$ and eventually saturates at $x_B = 1/3$. However, between this curve and the line $x_B = 1/2$, a lobe-shaped region develops within which $(\phi_0,\phi_2)$ is also an yrast state. Moreover, the upper and lower boundaries of this lobe approach $x_B = 1/2$ and $x_B = 1/3$, respectively, in the limit of large $\gamma$. In other words, the horizontal lines $x_B = 1/3$ and $x_B = 1/2$ act as asymptotes for the $K = 2$ critical curve. We therefore conclude that for $\kappa > 0$, the plane-wave state $(\phi_0,\phi_2)$ eventually becomes an yrast state at sufficiently large $\gamma$ for all values of $x_B$, except precisely at $x_B = 1/3$ and $x_B = 1/2$.

To understand this phenomenon, we revisit the analysis at the beginning of this section, where the energies of the plane-wave state $(\phi_0,\phi_k)$ and another state $(\psi_A,\psi_B)$ within the same $l = k x_B$ manifold are compared. If $(\psi_A,\psi_B)$ is a soliton state for which $|\delta \rho_A| \neq 0$ and $|\delta \rho_B| \neq 0$, the first term of $\delta \bar E_{\rm int}(\kappa > 0)$ in Eq.~(\ref{deltaEint}) can be made arbitrarily large by increasing $\gamma$. Consequently, even if $(\phi_0,\phi_k)$ is not an yrast state for $\kappa = 0$, its energy at $\kappa > 0$ can become lower than that of any soliton state in the $l = k x_B$ manifold for sufficiently large $\gamma$, irrespective of the value of $x_B$. However, this observation alone does not guarantee that $(\phi_0,\phi_k)$ will become an yrast state, since there may exist other plane-wave states within the same $l = k x_B$ manifold with even lower energies. To assess this possibility, we note that the minority concentration can in general be expressed as
\begin{align}
x_B = \frac{p}{q},
\end{align}
where $p$ and $q$ are coprime integers and $p\le q/2$. In the $l = k x_B$ manifold, there exist infinitely many plane-wave states of the form
\begin{align}
(\psi_A,\psi_B) = (\phi_{m p}, \phi_{k +mp- m q}),
\end{align}
with $m = 0, \pm 1, \pm 2, \ldots$, all of which share the same interaction energy. These states are distinguished by their kinetic energies,
\begin{align}
\bar E_{\rm kin} = l^2 + x_A x_B (k - m q)^2
\end{align}
and only the plane-wave state with the lowest kinetic energy can potentially serve as the yrast state. Since $k$ is restricted to positive integers, we require $(k-mq)^2  \ge k^2$ for all $m \ne 0$ for $(\phi_0,\phi_k)$ to be a viable candidate for the yrast state. Thus, $k$ must lie within the range
\begin{align}
k \le \left\lfloor \frac{q}{2} \right\rfloor,
\label{krange}
\end{align}
where $\lfloor q/2 \rfloor$ denotes the integer part (floor) of $q/2$. Once $k$ lies within this range, then the previous argument implies that $(\phi_0,\phi_k)$ will eventually become an yrast state for sufficiently large $\gamma$. Consequently, for a fixed $x_B = p/q$, the plane-wave states $(\phi_0,\phi_k)$ emerge as yrast states one by one with increasing $\gamma$, in order of increasing $k$, up to $k = \lfloor q/2 \rfloor$. Conversely, Eq.~(\ref{krange}) also means that the plane-wave state $(\phi_0,\phi_k)$ can never become an yrast state for those concentrations $x_B = p/q$ satisfying $\lfloor q/2 \rfloor < k$, regardless of how large $\gamma$ is. This implies, for example, that $(\phi_0,\phi_2)$ can never become an yrast state at $x_B = 1/2$ or $x_B = 1/3$, which is precisely why the critical curve $x_B(\gamma,K = 2)$ exhibits asymptotes at these values of $x_B$. Likewise, $(\phi_0,\phi_3)$ can never become an yrast state for $x_B = 1/2, 1/3, 1/4, 1/5,$ and $2/5$. This conclusion is confirmed by our numerical determination of the critical curve $x_B(\gamma, K = 3)$, shown along with a partial phase diagram in Fig.~\ref{cckp}, where these values indeed serve as asymptotes. The asymptotes of the critical curves for higher $k$ values can be deduced similarly. Once again, as in the case of $\kappa < 0$, the phase diagram shown in Fig.~\ref{cckp} provides a clear explanation for the nonanalytic structure of the yrast spectrum displayed in Fig.~\ref{yrastkp}.

\begin{center}
\begin{figure}[t]
     \includegraphics[width=8.6cm]{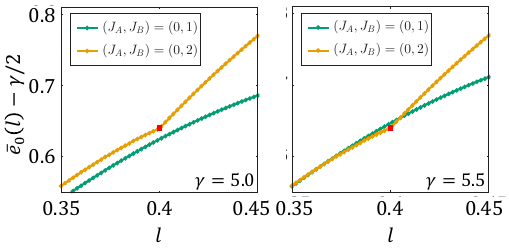}
\caption{The two lowest spectral branches in the vicinity of $l = 2x_B$ with $x_B = 0.2$, distinguished by the different winding numbers of the condensate wave functions, are shown for $\gamma = 5$ (left) and $\gamma = 5.5$ (right). At $\gamma = 5$, the branch containing a soliton state at $l = 2x_B$ constitutes the yrast spectrum, as it lies below the branch containing the plane-wave state $(\phi_0,\phi_2)$ (marked by the red square). As $\gamma$ increases, the plane-wave branch shifts downward relative to the soliton branch, and the plane-wave state $(\phi_0,\phi_2)$ becomes degenerate with the soliton state in the other branch at the critical value $\gamma_c \simeq 5.2$. The crossing of the two branches is clearly visible in the right panel at $\gamma = 5.5$.} 
\label{branchcross}
\end{figure}
\end{center}
Importantly, the critical curves determined here deviate significantly from those obtained perturbatively in Ref.~\cite{Wu2015}. To illustrate this, we focus on the $K=2$ critical curve. In Fig.~\ref{cccomparison}, we show the comparison of this critical curve determined from GP analysis and the perturbative approach in Ref.~\cite{Wu2015} for various $\kappa$ values.  We see that the perturbative approach only provides an accurate determination of the critical curve up to a certain $\gamma$ (or $x_B$) value  beyond which significant discrepancies from the exact GP analysis occur. Furthermore, such discrepancies occur at progressively smaller $\gamma$ and $x_B$ values as $\kappa$ increases, implying that the perturbative approach ceases to be valid as the degree of asymmetry increases.  Because the perturbative approach assumes a continuous transformation from soliton states to plane-wave yrast states,  this  indicates that such a physical scenario is no longer correct for sufficiently large $\gamma$ and $x_B$  when $\kappa > 0$. To substantiate this observation, we examine the evolution of the condensate wave functions at $l = 2 x_B$ for $x_B = 0.2$ and $\kappa = 1$ as $\gamma$ increases across the critical curve (see the dashed line with arrow in Fig.~\ref{cckp}). Judging from Fig.~\ref{cccomparison} we know that the perturbative approach is no longer valid for such a value of $x_B$.  In Fig.~\ref{wf_kp}, we plot the condensate wave functions at $\gamma = 3,4,5$ and $\gamma = 5.5$, corresponding to values just below and above the critical $\gamma$ value of $\gamma_c \simeq 5.2$. We observe that the condensate wave functions at $\gamma = 3,4$ and $5$ are soliton states that do not continuously evolve into the plane-wave state $(\phi_0,\phi_2)$ found at $\gamma = 5.5$. This indicates that the emergence of the plane-wave yrast state occurs via a branch crossing rather than through a continuous transformation. This process is illustrated explicitly in Fig.~\ref{branchcross}, where the lowest two branches of the energy spectrum, one containing the plane-wave state $(\phi_0,\phi_2)$ at $l = 2x_B$ and the other a soliton state, cross as $\gamma$ is varied between $5$ and $5.5$. 

\begin{center}
\begin{figure}[t]
     \includegraphics[width=8.6cm]{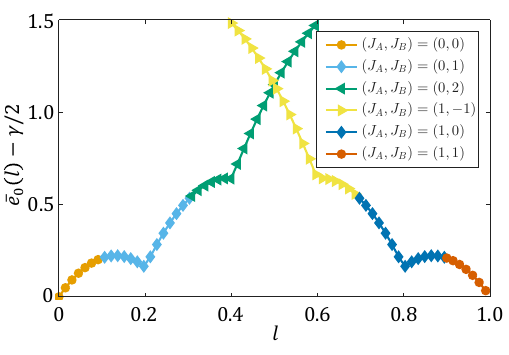}
\caption{The yrast spectrum plotted over the full range of $0<l<1$ for the system with $x_B = 0.2$, $\kappa = 1$ and $\gamma = 20$. The winding numbers of the condensate wave functions are explicitly shown for each section of the spectrum. } 
\label{branchcross2}
\end{figure}
\end{center}

\begin{center}
\begin{figure}[t]
     \includegraphics[width=8.6cm]{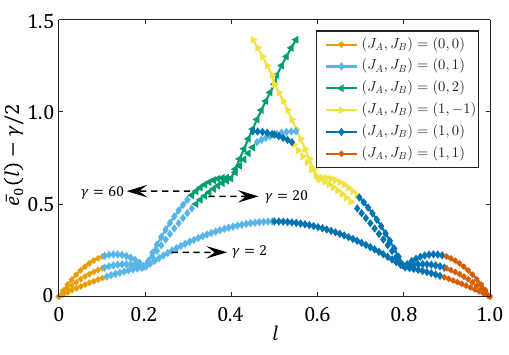}
\caption{The yrast spectra  for the system with $x_B = 0.2$ and $\kappa = 0.3$ at $\gamma = 2, 20$ and $60$.} 
\label{ffig}
\end{figure}
\end{center}

Now we make two observations on the derivative discontinuity of the spectrum at $l=0.5$ shown in Fig.~\ref{yrastkp}. First, such a discontinuity arises also from a branch crossing, except that this time both branches are solitons with different sets of winding numbers. This can be clearly seen when the spectrum is plotted over the full range of $0<l<1$ and with the winding numbers of each section explicitly displayed. Such a plot is shown in Fig.~\ref{branchcross2} for the system parameters $x_B = 0.2$, $\kappa = 1$ and $\gamma = 20$ (corresponding to one of the spectra in Fig.~\ref{yrastkp}). We see that there are two distinct soliton branches with winding numbers $(J_A,J_B) = (0,2)$ and $(J_A,J_B) = (1,-1)$ which cross at $l = 0.5$, implying that these distinct soliton solutions are degenerate at this value of angular momentum. The yrast spectrum, defined by the lowest energy solution, thus switches from one branch to another at $l = 0.5$, resulting in the observed derivative discontinuity of the spectrum. Second, such a discontinuity is not a universal feature for $\kappa >0$ but depends on the system parameters. We demonstrate this by plotting in Fig.~\ref{ffig} the yrast spectra  for the system with $x_B = 0.2$ and $\kappa = 0.3$ at $\gamma = 2, 20$ and $60$. We observe that the yrast spectrum is smooth at $l = 0.5$ for small values of $\gamma$ and that the derivative discontinuity emerges only when $\gamma$ becomes sufficiently large.

{Finally, we close this section by identifying a suitable experimental system and estimating the physical parameters relevant to the phase diagrams. Such a system has already been employed in experimental studies of persistent currents~\cite{Beattie13}: a toroidal two-component gas of $^{87}$Rb atoms occupying two different spin states. The gas is first confined to the horizontal plane by a harmonic trap of frequency $\omega_\perp$ and  then further confined within that plane to a ring of radius $R$. Denoting the inter-species s-wave scattering length by $a_s$,  the dimensionless  inter-component interaction strength can be written as
\begin{align}
\gamma = \frac{2NMRa_s\omega_\perp}{\pi \hbar }.
\end{align}
With typical experimental parameters: $N = 10^4$, $R = 10\mu$m, $\omega_\perp/2\pi = 350$Hz, and $a_s = 100 a_0$ ($a_0$ is the Bohr radius), we find that $\gamma \approx 1.1\times 10^3$. Since $a_s$ can be tuned by Feshbach resonance, the various regimes shown in the phase diagrams can be readily accessed with such an experimental system. }


\section{Conclusions}
In this paper, we have systematically determined the critical curves governing the existence of plane-wave yrast states in an asymmetric two-component Bose gas confined to a ring, using numerical solutions of the coupled GP equations supplemented by analytic insights. Because the emergence of these plane-wave yrast states gives rise to derivative discontinuities in the yrast spectrum, the resulting critical curves define a phase diagram that provides a clear picture of the analytic structure of the yrast spectrum. 

We find that the behavior of these critical curves depends crucially on the parameter $\kappa$, which characterizes the interaction asymmetry. For $\kappa < 0$, \emph{i.e.}, when the intra-component interaction is weaker than the inter-component interaction, the regions in the $\gamma$--$x_B$ plane that support various plane-wave yrast states are reduced relative to those of the SU(2)-symmetric system. Conversely, these regions are significantly enlarged in the $\kappa > 0$ regime. In fact, for sufficiently large $\gamma$, the plane-wave state $(\phi_0,\phi_k)$ can become an yrast state for all but a discrete set of $x_B$ values. These excluded values can be identified using simple energetic considerations. Furthermore, the mechanism by which plane-wave yrast states emerge as the interaction strength $\gamma$ varies depends sensitively on the interaction asymmetry. For $\kappa < 0$, they appear as the endpoint of a continuous evolution of soliton states at the same angular momentum, whereas for $\kappa > 0$ the situation is more complex. For sufficiently small values of $x_B$ the continuous evolution remains in place but for larger values of $x_B$ the plane-wave yrast states emerge by overtaking soliton states through a branch crossing. This explains why the critical curves established by the perturbative approach, which assumes a continuous evolution, is no longer accurate for  $\kappa>0$.
Our results demonstrate that interaction asymmetry plays a central role in shaping the rich and intricate analytic structure of the yrast spectrum in two-component Bose gases, a prediction that may in principle be tested experimentally.

\label{Conclusions}

\section*{Acknowledgements}  

 This work is supported by National Key R$\&$D Program of China (Grant No. 2022YFA1404103), Natural Science Foundation of China (Grant No.~12474264), Guangdong Provincial Quantum Science Strategic Initiative (Grant No.~GDZX2404007). S.Z. acknowledges support from HK GRF (Grant No. 17306024), CRF (Grants No. C6009-20G, No. C7012-21G, No. C4050-23GF), a RGC Fellowship Award No. HKU RFS2223-7S03 and Guangdong Provincial Quantum Science Strategic Initiative (Grant No.~GDZX2404001).

\bibliography{ref}	

 \end{document}